# Elliptic Vortices in Composite Mathieu Lattices

Fangwei Ye[1], Dumitru Mihalache[2] and Bambi Hu[1,3]

[1]*Department of Physics, Centre for Nonlinear Studies,*

*and The Beijing-Hong Kong-Singapore Joint Centre*

*for Nonlinear and Complex System(Hong Kong),*

*Hong Kong Baptist University, Kowloon Tong, China*

[2] *Horia Hulubei National Institute for Physics and Nuclear Engineering (IFIN-HH),*

*Department of Theoretical Physics, 407 Atomistilor, Magurele-Bucharest, 077125, Romania*

[3]*Department of Physics, University of Houston, Houston, Texas 77204-5005, USA*

We address the elliptically shaped vortex solitons in defocusing nonlinear media imprinted with a composite Mathieu lattice. Elliptic vortices feature anisotropic patterns both in intensity and phase, and can only exist when their energy flow exceed some certain threshold. Single-charged elliptic vortices are found to arise via bifurcation from dipole modes, which, to the best of our knowledge is the first example in the context of optics studies of symmetry breaking bifurcations for the phase dislocations of different dimensionalities. Higher-order elliptic vortices with topological charge *S* could exhibit spatially separated *S* single-charged phase singularities, leading to their stabilization. The salient features of reported elliptic vortices qualitatively hold for other elliptic shaped confining potentials.

# 1. Introduction

Spatial, temporal and spatiotemporal optical solitons are localized electromagnetics wave-packets, which can form in a variety of media characterized by different kinds of optical nonlinearities, such as cubic (Kerr-like), saturable, thermal, photorefractive, and quadratic ones [1-6]. In the past years there has been a huge interest in the theoretical and experimental study of multidimensional localized structures of different topologies forming both in optics [5] and in Bose-Einstein condensates (BECs) [7]. In optics, these localized (soliton-like) structures, either nondissipative or dissipative [8, 9] ones are quite complex physical objects and they represent the "particle-like" counterpart of the more common extended light structures.

Solitons with elliptic shapes naturally do not exist in isotropic media, unless their intrinsic anisotropic diffraction is compensated by other counteracts, such as anisotropic confining potential, correlations statistics [10,11], or the inherent anisotropy that the nonlinear material itself has [12,13]. The simplest elliptic solitons are bright spots with different widths in the two orthogonal directions, and they are realized in optics as incoherent excitations in isotropic nonlinear media [10,11] and also observed as coherent states in nonlocal nonlinear media [14,15]. Non-circular vortices and other more complex topological structures are of much interest in the context of BEC due to the experimentally readily achieved anisotropic harmonic trapping. In this respect, in rotating BEC systems, the role of asymmetries and confining potential ellipticities on vortices was first theoretically considered in Ref. [16] and further developed in Ref. [17]. The role of ellipticity on the stability of vortices in non-rotating traps was investigated in Ref. [18], where stable asymmetric vortices were reported to form for large enough nonlinearities, a result that is also found in the present work, though the confining external potential has a quite different origin, *viz*., the composite Mathieu optical lattice is realized by the superposition of two Mathieu beams with the same order and opposite parities. Soliton dipoles, vortex-antivortex pairs ("vortex dipoles") and vortex clusters in asymmetric nonrotating confining potentials (asymmetric traps) were also investigated [19-24]. Vortex dipoles, i.e., composite states with two embedded vortices of opposite topological charges (vortex-antivortex pairs), do exist as stationary soliton solutions to the effectively two-dimensional Gross-Pitaevskii equation for both symmetric and asymmetric traps. However, it should be mentioned that in circularly symmetric traps, vortex dipoles cannot exist in the linear limit and they bifurcate from soliton dipoles [20] or from dark soliton-like modes [25]. However, in asymmetric traps, vortex dipoles exist even in the linear limit [20]. Though the role of ellipticities on the formation and stability of vortices and other topological structures has been thoroughly studied in the context of BEC, there are very few works investigating the elliptic topological solitons in the context of optics. This is due to a lack of suitable media for their formation as well as their high chances of instabilities [12, 26]. The only stable elliptic optical vortices so far reported are encountered in nonlocal nonlinear media with the rotation as a necessary condition for their existence and stabilization [27].

The shapes and properties of solitons greatly depend on the topology of the underlying optical lattices. The Mathieu lattice [28,29], the third type of non-diffracting optical lattices after $\cos^2$ lattice and the Bessel lattice [30], features elliptic-shaped ring structures with its ellipticity accounted for by a parameter $e$. In the limit $e \to 0$, the Mathieu lattice returns back to the circular Bessel lattice, which is found to support stable ring vortices in defocusing nonlinear media [30]. One therefore will naturally ask: Does the Mathieu lattice support elliptic-shaped vortices? If the answer is yes, how do they differ from their circular counterparts in the case of Bessel lattice? We note that there is an important difference between the Bessel lattice and the Mathieu (or other elliptic) lattice: in the linear regime, a circular lattice itself supports a vortex mode, while an elliptic lattice does not. So, the existence of vortex modes in elliptic lattices is a nontrivial problem.

In this paper, using the Mathieu lattice as a physically feasible elliptic lattice, we find that the elliptic vortices exist if the energy flow exceeds a certain minimum value, in contrast with the circular vortices which have also their linear counterparts, i. e., they exist for any values of the energy flow, including the limiting case of arbitrarily small energy flows. We reveal that the origin of elliptic vortices is a result of a symmetry breaking bifurcation from dipole solitons, a nontrivial bifurcation which has not been reported before in the context of optics, to the best of authors' knowledge. The bifurcation of vortices from dipoles accounts for the asymmetric field profiles observed in elliptic vortices. We also find that the higher-order elliptic vortices are characterized by a spatial separation of several single-charged phase singularities. The possibility of finding stable multiply charged vortices whose embedded phase singularity is unfolded constitutes a central result of this work, which has not been revealed before to the best of our knowledge.

## 2. Model

We start the analysis with a nonlinear Schrödinger equation for the dimensionless complex amplitude for the electromagnetic field $q$ describing the light propagation along the $z$ axis in a bulk medium with a defocusing Kerr nonlinearity and a transverse modulation of refractive index

$$i\frac{\partial q}{\partial z} = -\frac{1}{2}(\frac{\partial^2 q}{\partial x^2}+\frac{\partial^2 q}{\partial y^2}) + |q|^2 q - p|R(x,y)|^2 q, \qquad (1)$$

Here the longitudinal $z$ and transverse $x$, $y$ coordinates are scaled to the characteristic diffraction length and beam width, respectively. $p$ stands for lattice depth, and the function $R(x,y)$ ($\max[R(x,y)]=1$) describes the lattice profile assumed to be induced optically by a superposition of two Mathieu beams with the same order and opposite parities (thus the name of the composite Mathieu lattice). Mathematically, this is written by a Whittaker integral [29, 31],

$$R(x,y) = \int_0^{2\pi} [Ace_m(\mathbf{n};e) + iBse_m(\mathbf{n};e)]\exp[ik_t(x\cos\mathbf{n} + y\sin\mathbf{n})]d\mathbf{n}, \qquad (2)$$

with $ce_m(\mathbf{n};e)$ and $se_m(\mathbf{n};e)$ being the $m$th-order even and odd angular Mathieu functions. The normalization coefficients $A$ and $B$ are used to ensure the same power of the even and odd components in Eq. (2). Here $e$ characterizes the ellipticity of the lattice. In the limit $e \to 0$, one recovers the radially symmetric Bessel lattice. An increase of $e$ from zero results in the stretching of the whole structure in the horizontal direction, thus the lattice displays a set of confocal elliptic rings (see Fig.1 (a)). We note that when $e$ exceeds some critical value $e_c$, the lattice will deviate itself from the elliptic structures since the Mathieu lattice transforms into a quasi-one-dimensional lattice when $e \to \infty$; we therefore limit the ellipticity to the interval $e \leq e_c$, for which the lattice displays a well-shaped elliptic structure. Further, higher $m$ (lattice order) has a larger $e_c$, therefore, to allow the appearance of a significant elliptic structure, we fix $m = 5$. In this case, $e_c \approx 10$. Experimentally, such lattice-creating Mathieu beams could be generated by holographic techniques [32], or by illuminating a narrow annular slit with a Gaussian aperture placed in the focal plane of a lens [33]. Equation (1) might find application in some photorefractive semiconductor crystals belonging to the $\bar{4}3m$ point symmetry group, with the polarization of soliton and lattice-creating beam orthogonal to each other [34]. Equation (1) also holds for lattice-trapped BECs with repulsive inter-atomic interactions. Eq. (1) conserves the energy flow $U = \iint |q|^2 dxdy$. Without loss of generality, $k_t = 2$ is used below.

We searched for vortex solutions of Eq. (1) in the form of $q(x,y,z) = (u_r + iu_i)\exp(i l z)$. Here $u_{r,i}(x,y)$ are real functions independent of the propagation variable $z$, while $l$ is the propagation constant. Substitution into Eq. (1) yields a two-dimensional nonlinear eigenvalue problem which was solved by a Newton relaxation method [35]. In all cases analyzed in this work, the chosed numerical method converged, after several iterations, to a stationary solution up to a prescribed accuracy $[O(10^{-8})]$, provided that a proper guess for the initial field distribution was selected. The initial field distributions were chosen as the circularly symmetric Laguerre-Gaussian modes which are characterized by a nonzero integer standing for the topological charge.

## 3. Single-charged elliptic vortices

A typical example of the amplitude and phase of single-charged vortices is shown in Fig. 2(a). The elliptic vortex shows an azimuthally modulated intensity profile whose maximum and minimum are achieved respectively, at major and minor semi-axes of the ellipse. The phase gradient is also not uniform. Although the phase change along a closed trajectory containing the phase singularity is $2p$ (for the single-charged vortex), the phase varies fastest near the minor axis, while slowest at the major one. With the increase of the lattice ellipticity, such anisotropy gets more pronounced.

As mentioned above, the composite Mathieu lattice itself does not support any type of vortex modes, a fact that can be readily confirmed by solving the eigen-equation of the linear version of Eq. (1). Thus, unlike their circular counterparts in the case of Bessel lattice, elliptic vortices exist only when their energy flows exceed some certain threshold s; see Fig. 3(a), which shows the energy flow $U$ versus the propagation constant $l$ for vortices of different ellipticities. Along with vortices, we also plot in Fig. 3(a) the $U = U(l)$ curve for dipole solitons. Dipole solitons appear as two "crescent"-like bright intensity spots with $\pi$ phase jump between them (see Fig.2 (c)). As Fig. 3(a) shows, for Bessel lattices ($e=0$), both vortices and dipoles exist in the linear limit (i. e., for any values of $U$ including the case of arbitrarily small values of the energy flow). However, for $e \neq 0$, vortex solitons exhibit a threshold $U_{cr}$ (and correspondingly, $\mathbf{l}_{cr}$), below which they cease to exist. In contrast, dipole solitons with the orientation of their nodal lines at the minor axis remain thresholdless due to the existence of linear dipole modes in elliptic lattices. Importantly, the $U = U(\mathbf{l})$ diagram shows that the elliptic vortices bifurcate from dipole modes upon increase of the energy flow. Such bifurcation may seem surprising since "conventional" vortices and dipoles have essentially different phase patterns: the former are recognized by a screw phase structure whereas the latter by a $\pi$ phase jump. Nevertheless, by following the phase pattern changes of the elliptic vortices on the $U=U(\mathbf{l})$ diagram, one finds the bifurcation of vortices from dipole modes possible in the case of elliptic vortices. At high energy flows, i. e., far from the bifurcation point $U_{cr}(\mathbf{l}_{cr})$, vortices and dipoles are quite different both in their intensity and phase pattern (cf. Fig. 2 (a) and Fig. 2(c)). On the other hand, with the decrease of $U$, the nonlinear effect decreases while the linear lattice effect (which is anisotropic) gets more and more significantly in determining the vortex structure. Finally, near the cutoff (bifurcation point), the vortex even evolves into two weakly interacting "crescent"-like spots, whose phase difference turns out to be almost $\pi$ (see Fig. 2(b)). In another words, as a consequence of its asymmetric structure, the vortex near the cutoff resemble itself to be a dipole state (cf. Fig. 2 (b) and Fig. 2(c)). Therefore, the elliptic vortex branches out from the dipole mode via a symmetry breaking bifurcation. The quantitative value of the energy flow $U_{cr}(\mathbf{l}_{cr})$ at the bifurcation point is observed to increase with the increase of lattice ellipticity (see Fig.3 (b)).

The bifurcation of such localized vortices from dipole modes constitutes a nontrivial example of symmetry breaking bifurcation, which, to the authors' knowledge, has not been reported before in the context of optics. However, two examples of *one dimensional* version of such bifurcation are reported in the theory of nonlinear optical surface waves [36-38]. The first example provides an example of bifurcation of the asymmetric TE mode from the symmetric one in a symmetric layered structure, which is composed of a linear dielectric film bounded on both sides (the cladding and substrate media) by the same nonlinear Kerr medium [36, 37]. The second example is provided by a nonlinear Kerr dielectric slab sandwiched between two same metals, where the asymmetric TM surface plasmonic mode bifurcates from the symmetric one [38]. In both physical settings, the symmetric surface modes also exist in the linear regime, whereas the asymmetric modes bifurcating from the symmetric one exist only above a certain threshold energy flow. The symmetric modes and the asymmetric bifurcated modes are similar at the bifurcation point, while quite different far from the bifurcation point. With reference to our physical setting, one can see that the role of the "symmetric mode", which exists also in the linear limit, is played by the dipole modes, whereas the role of the bifurcated "asymmetric mode" is played by the vortex soliton (note that the vortex mode is asymmetric with respect to the elliptic lattice). To the authors' knowledge, this is the first example of topological complex modes in the form of vortices bifurcating from dipoles, which provides a smooth transition between screw dislocation and edge dislocation [39]. In view of the fact that the screw dislocation is a two-dimensional (2D) phase object while the edge dislocation is essentially a one-dimensional (1D) object, we emphasize that such bifurcation also bridges the phase dislocations of different dimensionalities, which is impossible to be explored in the framework of one-dimensional symmetry breaking bifurcations. The vortex originating via bifurcation from dipoles also accounts for the intensity modulations and the azimuthally varying phase gradients seen in the structure of elliptic vortices (see Fig. 2), as well as the occurrence of an energy flow threshold for their existences (see Fig. 3). It is worthy to notice that such bifurcation patterns are generic ones and they also occur in other physical settings, e. g., in the context of BEC it is possible the transformation of dark nodal solitons from one-dimensional stripes to two-dimensional vortex solitons when the transverse confinement exceeds a critical value [40]. Thus the dipole solitons existing also in the linear limit in the present setting play the role of dark solitons, which have linear counterparts too, in confined BECs. However, a distinct difference between our bifurcation and that in Ref. 40 is manifested in the different nature of soliton states undertaking the bifurcation. In Ref. 40, the bifurcation is carried on by dark solitons, which are delocalized states and thus strongly rely on the external parameters like the transverse confinement; in contrast, the soliton states (both dipoles and branched vortices) in our setting are localized entities and thus the bifurcation is a pure internal process realized by solitons themselves. As another typical example we mention the transformation of ring dark solitons into robust clusters of vortex pairs upon onset of the snake-type instability, a phenomenon which was put forward in Ref. 41.

## 4. Multiple-charged elliptic vortices

Higher-order vortices are also found to exist in elliptic lattices. Like the single-charged vortices, the multiple-charged vortices also feature azimuthally asymmetric phase gradient and intensity profiles. Figures 4(a) and 4(b) present a typical example of stable double-charged elliptic vortex. Remarkably, unlike the conventional double-charged vortex where the embedded phase singularity is double-folded, the elliptic vortex shows an unfolded behavior in the phase pattern, with the appearance of two single-charged phase singularities separated by a finite distance $d$. Separation $d$ is found to depend significantly on the lattice ellipticity, and for a fixed propagation constant, a higher ellipticity of the lattice leads to a larger value of $d$, while the zero-ellipticiy limit leads to a circular vortex where the two $S=1$ phase singularities merge into a single phase singularity with topological charge two. For a fixed ellipticity, we found that the separation distance $d$ is a monotonically increasing function of energy flow (see Fig. 4(c)). Interestingly enough, $d$ is numerically found to asymptotically approach $d_{I\to 0} = 2\sqrt{e}/k_t$ (i.e., half of the distance between the foci of the ellipse). Similar scenario for singularities unfolding is also found for higher-order vortices (see Fig. 5 for a typical example of triple-charged vortex). The unfolding of higher-order phase singularities of elliptic vortices has an essential impact on their stability, see below.

## 5. Linear stability analysis

In view of the highly asymmetric pattern of elliptic vortices, a crucial issue is their stability. We performed both a linear stability analysis and a direct integration of the governing equations (1) with the input condition $q|_{z=0} = (1+r)w$, where $r$ stands for the input noise with the variance $s^2 = 0.01$. For fundamental vortices, we find that they are stable in their whole existence domain provided that the lattice depth is large enough, which is similar to their counterparts in Bessel lattice. It is worthy to mention that, at the lattice order as high as $m=5$ as in the present study, all of the fundamental elliptic vortices were found to be stable irrespective of the lattice depth. For higher-order vortices, their stability exhibits novel characteristics associated with the phase singularity unfolding. At low energy flows where the separation distances between phase singularities are small, they are unstable, gradually evolving into ring structures with uniform phase distribution. However, with the increase of the energy flow, which is accompanied by an increase in the singularity separation, elliptic vortices turn out to become very stable entities. The stability and instability domain for charge two vortices are depicted in Fig.4(c), for the lattice depth $p=10$. Note that such stability characteristics remain at higher lattice depths: the higher-order vortices are unstable at low energy flows and are stable for energy flows above certain threshold values, or correspondingly, above certain value of the phase singularity separations (Fig.4(c)). Thus, the stability properties of elliptic vortices are significantly different to those corresponding to the case of Bessel lattice, where the instability can only be found in the high energy flow regime while the stability could appear in the low energy flow regime. A typical example of stable propagation of a triple-charged vortex in the presence of input random noise is shown in Fig. 5. It is clearly seen from this figure that the relatively strong white noise added to the input vortex is cleaned out after propagation and that the multiple-charged vortex maintains its intrinsic structure upon propagation in a robust way, both in amplitude and phase.

## 6. Discussion and conclusions

Before concluding, we emphasize that, although the elliptic vortices presented in this work were investigated in the framework of the composite Mathieu lattice which features a particular modulation, the reported results are found to qualitatively hold true for other types of elliptic confining potentials, e.g., for the elliptic potential with uniform intensity in the direction of the angular elliptic variable (see Fig. 1 (b)). In this respect, we bring the reader's attention again onto Ref. [18] where another quite different confining elliptic (asymmetric) potential, i.e., a parabolic trapping potential with different confining strengths in its two transverse directions is also found to support stable elliptic vortices for large enough nonlinearities. However, we have further revealed in the present study that the emergence of the elliptic vortices in generic asymmetric confining potentials is associated with the occurrence of a symmetry breaking bifurcation from soliton dipoles, if the energy flow is larger that a certain threshold value.

It is worthy to note that although the symmetry braking bifurcations involving somehow complex topological structures such as vortex-antivortex pairs (vortex dipoles) have been revealed in a couple of previous works [20,25], the corresponding bifurcation associated with simpler vortex structures such as the single ring-like vortex has not been addressed before to the best of our knowledge. Note also an essential difference between the symmetry breaking bifurcation revealed in the present work and the bifurcation studied in Refs. 20 and 25: in those works the branched vortex dipole has the overall topological charge equal to zero, a value which is identical to that of the soliton from which it branches out. In contrast, we found that in elliptic traps stable vortices with nonzero topological charges could emerge from vorticityless dipoles providing a smooth transition between edge and screw dislocations (recall that the bifurcations studied in Refs. 20 and 25 were put forward in BECs confined in radially symmetric traps).

In conclusion, we have revealed the salient features of elliptic vortices supported by an elliptic confining potential in defocusing nonlinear media. They exist when the energy flow exceeds a certain threshold, and feature anisotropic field distribution. Fundamental elliptic vortices emerge via symmetry breaking bifurcation from dipole modes, a nontrivial example of bifurcation that provides a smooth transition of 2D screw dislocations from 1D edge dislocations. The emergence of vortices from dipoles also accounts for the anisotropic field pattern observed in elliptic vortices, as well as for the existence of a minimum energy flow threshold for their formation.

We have also studied the influence of the ellipticity of the confining potentials on the higher-order charged vortices. We have revealed the existence of higher-order stable vortices that in symmetric potentials cannot even exist [42]. The striking finding is that their phase singularities may unfold into several spatially separated single charge phase singularities, which greatly impact their stability properties.

F.Ye thanks Liangwei Dong and Dahai He for helpful discussions. This work is supported in part by Hong Kong Baptist University and the Hong Kong Research Grants Council.

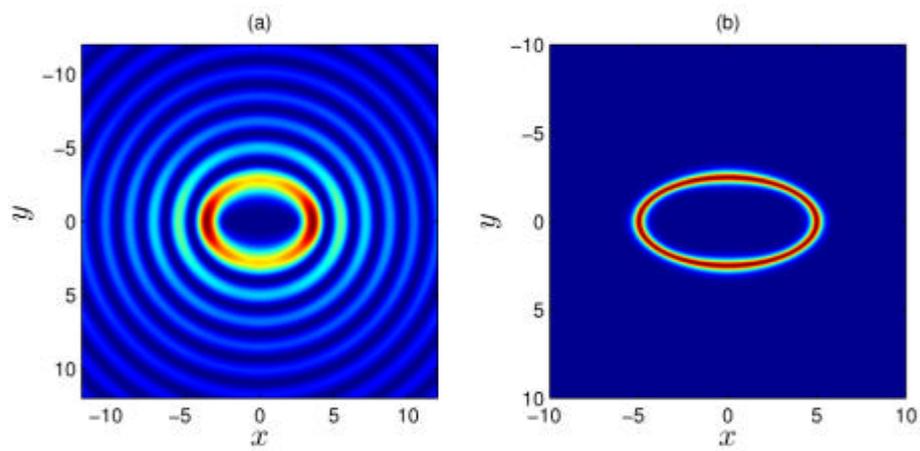

Figure 1. (Color online) (a) Composite Mathieu lattice with $m=5, e=5, k_t = 2$. (b) Elliptic ring lattice without intensity modulation in the direction of angular elliptic variable. Semi-major axis $a=5$ and semi-minor axis $b=2.5$. All quantities are plotted in arbitrary dimensionless units.

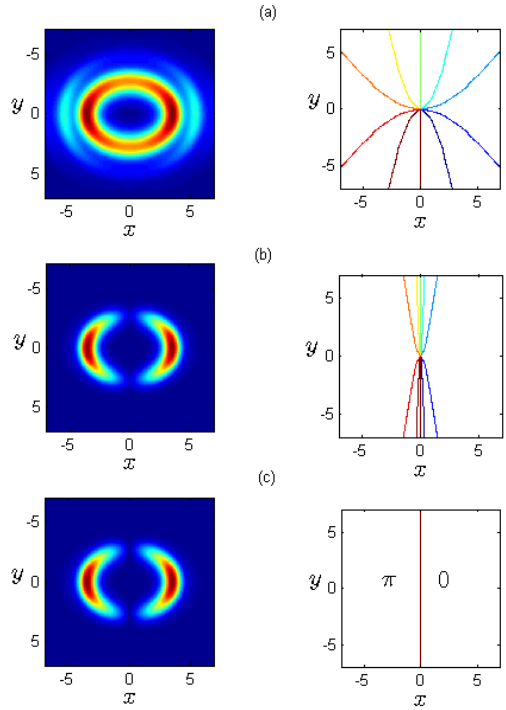

Figure 2. (Color online) Amplitude (left column) and contour plots of phase (right column) for single-charged vortices at (a) $l = 2$, (b) $l = 4.6$ and for dipole soliton (c) at $l = 4.6$, respectively. Contour lines are spaced $p/5$ units. The $p$ phase difference between two constituting bright spots of the dipole (c) is indicated. In all plots, $m = 5$, e=5, p=10. All quantities are plotted in arbitrary dimensionless units.

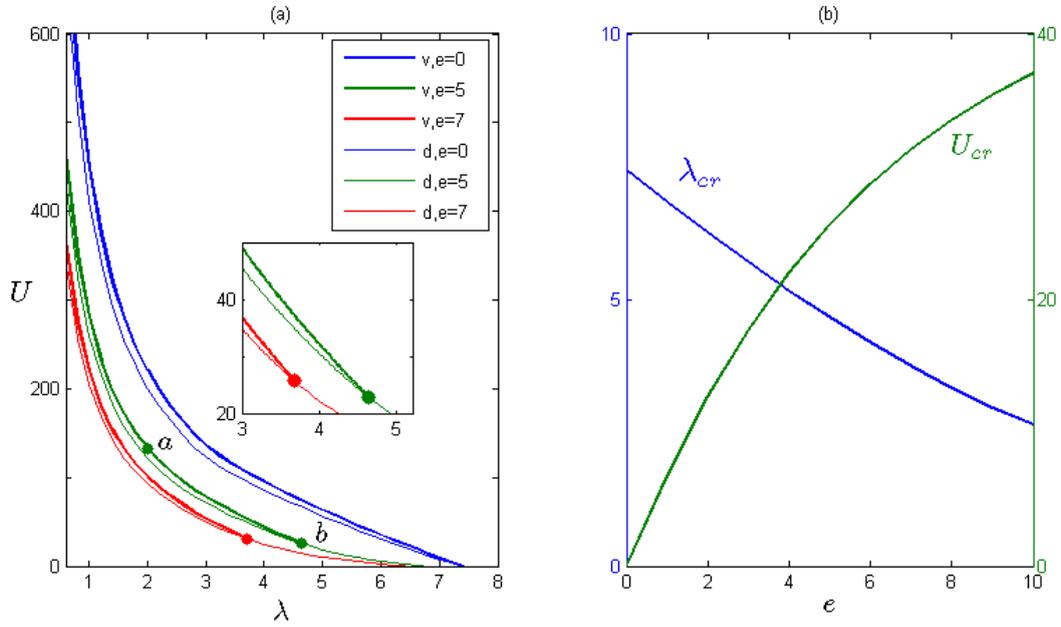

Figure 3. (Color online) (a) Energy flow versus propagation constant for vortices (thick lines) and dipoles (thin lines) at different lattice ellipticities (Inset: the curves near the bifurcation points). Points marked by *a* and *b* correspond to the elliptic vortices shown in Figs. 2(a) and 2(b) respectively. (b) Critical (threshold) propagation constant and energy flow versus lattice ellipticity. Here $m = 5$, p=10. All quantities are plotted in arbitrary dimensionless units.

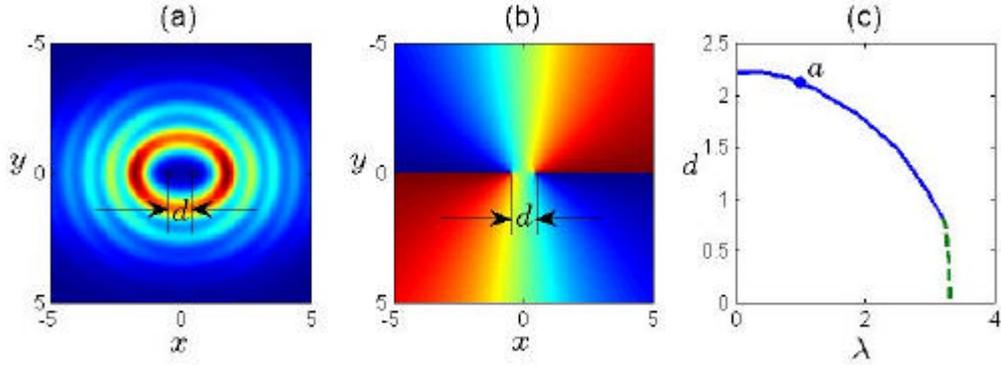

Figure 4. (Color online) (a) Amplitude and (b) phase of a double-charged vortex at $l = 1$ with its two phase singularities separated by a distance $d$. (c) Phase singularities separation versus propagation constant for double-charged vortices. Solid line corresponds to stable solutions while dashed to unstable solutions. Point $a$ on the curve corresponds to the vortex shown in Fig. 4 (a) and (b). Here $m = 5$, e=5, p=10. All quantities are plotted in arbitrary dimensionless units.

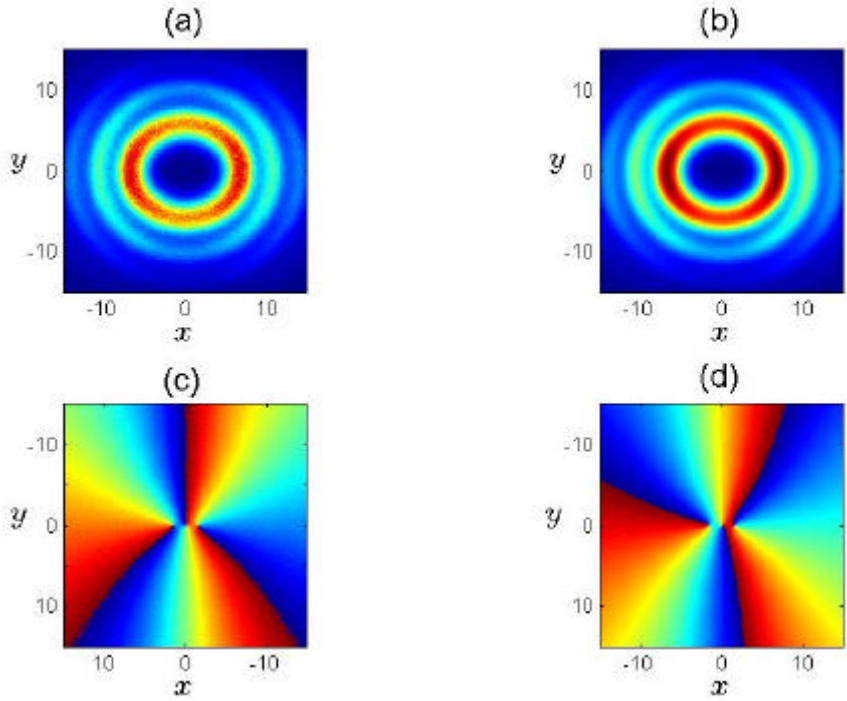

Figure 5. (Color online)Stable propagation of a triple-charged vortex. Left panel: distributions of (a) amplitude and (c) phase at $z = 0$; Right panel: distributions of (b) amplitude and (d) phase at $z = 300$. White noise with variance $s^2 = 0.01$ was added to the input field distributions (left panel). Here $l = 1.5, m = 5$, e=5, p=10. All quantities are plotted in arbitrary dimensionless units.